\documentclass[printer]{aa}

\usepackage{natbib}
\usepackage{amsmath}
\usepackage{graphicx}
\usepackage{txfonts}
\usepackage{color}

\begin{document}
\title{Coupled jet-disk model for Sgr A*: explaining the flat-spectrum radio core with GRMHD simulations of jets}
\titlerunning{On the flat-spectrum radio core in Sgr~A*}

\author{Monika Mo{\'s}cibrodzka~\inst{1} \and Heino Falcke~\inst{1,2}}
\institute{$^1$Department of Astrophysics/IMAPP,Radboud University
  Nijmegen,P.O. Box 9010, 6500 GL Nijmegen, The Netherlands\\
$^2$ASTRON, Dwingeloo, The Netherlands\\
\email{m.moscibrodzka@astro.ru.nl}, \email{h.falcke@astro.ru.nl}}
\date{Received 17 September 2013; accepted 16 October 2013}
\renewcommand{\labelenumi}{\alph{enumi})}

\newcommand{\rg}{GM_{BH}/c^2}
\newcommand{\Rg}{R_g}
\newcommand{\trat}{T_{\rm p}/T_{\rm e}}
\newcommand{\Te}{T_{\rm e}}
\newcommand{\Tp}{T_{\rm p}}
\newcommand{\mdot}{\dot{M}}
\newcommand{\mdotu}{\rm M_\sun yr^{-1}}

\newcommand{\shortened}[1]{\textbf{\it#1}}
\newcommand{\revised}[1]{\textbf{#1}}

\abstract
{The supermassive black hole in the center of the Milky Way, Sgr~A*, displays
  a nearly flat radio spectrum that is typical for jets in active galactic
  nuclei. Indeed, time-dependent magnetized models of radiatively
  inefficient accretion flows (RIAFs), which are commonly used to explain the
  millimeter, near-infrared, and X-ray emission of Sgr A*, often also produce
  jet-like outflows. However, the emission from these models has so far failed
  to reproduce the flat radio spectrum.}
{We investigate whether current accretion simulations can
  produce the compact flat spectrum emission by simply using a
  different prescription for the heating of the radiating particles in the jet.}
{We studied the radiative properties of accretion flows
  onto a black hole produced in time-dependent general-relativistic
  magnetohydrodynamic (GRMHD) simulations. A crucial free parameter in
  these models has always been the electron temperature, and here we
  allowed for variations in the proton-to-electron temperature ratios in the
  jet and disk.}
{We found that the flat spectrum is readily reproduced by a standard
  GRMHD model if one has an almost isothermal jet coupled to a
  two-temperature accretion flow.  The low-frequency
  radio emission comes from the outflowing sheath of matter
  surrounding the strongly magnetized nearly empty jet. The
  model is consistent with the radio sizes and spectrum of Sgr~A*.}
{Hence, GRMHD models of accreting black holes can in principle
  naturally reproduce jets that match observed characteristics. For
  Sgr A* the model fit to the spectrum predicts higher mass-accretion
  rates when a jet is included than without a jet. Hence, the impact of the
  recently discovered G2 cloud that is expected to be accreted onto Sgr A*
  might be less severe than currently thought.
}
  
\maketitle
\section{Introduction}

 \citet{blandford:1979} suggested that the flat-spectrum
  radio cores seen in many active galactic nuclei are the optically
  thick parts of conical plasma jets. Very Long Baseline
  Interferometry (VLBI) has indeed revealed a jet-like geometry in
  many flat-spectrum compact radio cores. Interestingly, the center of
  our Galaxy also hosts a flat-spectrum core, called Sgr A*, and it
  has been suggested that it may be associated with a relativistic
  downsized jet from a starving supermassive black hole (BH)
  (\citealt{falcke:1993,falcke:2000,markoff:2007}). Unfortunately,
  scattering by interstellar electrons smears out the source
  structure. Nonetheless, the presence of a relativistic outflow in
  Sgr~A* is strengthened by the size-frequency relation
  \citep{bower:2004} and by a 20-minute time-lag between flares in the 43
  and 22 GHz light curves \citep{yusef:2006}, which can be well explained
  by a jet model \citep{falcke:2009}.

 At millimeter wavelengths, the spectrum of Sgr A* peaks in the
  so-called submillimeter bump \citep{falcke:1998}, which can be
  modeled by synchrotron emission arising from a radiatively
  inefficient accretion flow onto a BH (RIAFs; first applied
  to Sgr~A* by \citealt{narayan:1995}). Over the past years, the RIAF
model has progressed from a simple semi-analytical model to more
complex time-dependent general-relativistic
  magnetohydrodynamic (GRMHD) models. In these models relativistic
jets are often produced \citep[e.g.,][]{beckwith:2008}. The
synchrotron emission from a GRMHD-RIAF can now be calculated with a
high degree of precision using general relativistic radiative-transfer
codes (e.g.,
\citealt{broderick:2009a,dolence:2009,dexter:2009,roman:2012}).
Advances in numerical modeling, together with less scattered, mm-wave
VLBI measurements of Sgr A*, provide an opportunity to put tight constraints
on jet-plus-disk models for BHs (\citealt{broderick:2009b,dexter:2012}).

So far, GRMHD-RIAF models have not naturally reproduced the radio
spectrum of Sgr A* --- with or without a jet. This may be partly
because of the uncertainty in how to treat plasma temperatures. The
dynamics of the plasma around the BH is sensitive to the
temperature of protons $\Tp$, whereas the radio synchrotron spectra
depend on the temperature of electrons $\Te$.  In a collisionless
plasma system, such as Sgr A*, the strength of $e$--$p$ coupling is
unknown. Modeling the effect of the $e$--$p$ energy exchange is very
simplified or not considered in typical MHD simulations. In a
standard approach one often assumes that $\trat=const$ in the entire
simulation domain (e.g., \citealt{moscibrodzka:2009,moscibrodzka:2012}
consider $\trat=1,3$, and 10). The constant $\trat$ everywhere
suppresses emission from a GRMHD outflow because the outflow from
near the event horizon naturally must have much lower densities than
the inflow.

In this work, we relax the assumption of constant
  $\trat$. Indeed, early RIAF disk models assumed a high $\trat$ to
account for the submillimeter bump, while early jet
  models assumed a rather high $\Te$ to reproduce the radio emission of
Sgr~A*. Only later $\trat$ was decreased in RIAF models to allow for
lower accretion rates imposed by Faraday rotation measurements
(\citealt{BowerFalckeWright2005a,marrone:2007}). 

\citet{yuan:2002} pointed out that the electron temperature in the
disk should be lower than that of the jet by a factor of ten to make
both disk and jet contribute to the emerging spectrum.  Here, we adopt
a similar approach. The use of a higher $\Te$ in the GRMHD jets is
motivated by the presence of several physical processes that are
enhanced in the outflows and may cause stronger heating, such as
stronger plasma magnetization, stronger shearing motion in the jet
sheath, and shocks. All these processes are observed in the
GRMHD simulations of jets (Brinkerink et al., in prep.).

The paper is organized as follows. In Sect.~\ref{model}, we briefly describe
the GRMHD model of the BH accretion flow, the radiative transfer
technique, and the electron temperature parameterization used to compute
spectra and images of the jet-plus-disk system. We present and discuss the new
results in Sects.~\ref{results} and~\ref{discussion},
respectively.

\begin{table}[t]
\caption{List of radiative-transfer models.$^*$preferred $\trat$ in the  disk.}
\label{tab:1}      
\centering                 
\begin{tabular}{c c c c c c c }   
\hline\hline               
 $i \mathrm{[\degr]}$&$\Theta_{\rm e,j}$ & $^*(\trat)_{\rm d}$ & $n_{\rm e,0} {\rm [cm^{-3}]}$& $B_{\rm 0} {\rm [kG]}$ & $\mdot [\mdotu]$\\
\hline                        
 $90$ & $30$ & $15-20$ & $7\times 10^8$ & $3.6$ &$10^{-7.3}$\\
 $60$ & $30$ & $15-20$ & $7\times 10^8$ & $3.6$ &$10^{-7.3}$\\
 $30$ & $30$ & $10-15$ & $4\times 10^8$ & $2.7$ &$10^{-7.6}$\\
\hline                     
\end{tabular}
\end{table}

\section{Model description}\label{model}

To investigate the radiative properties of the jet--disk--BH triad, 
we split the numerical modeling into three steps: 
(1) we computed the evolution of the GRMHD flow onto a BH;
(2) we rescaled the dynamical model to Sgr~A*; 
and (3) we computed synchrotron spectra and images of the system.

The accretion-flow evolution was calculated by using the axisymmetric
GRMHD code HARM-2D \citep{gammie:2003}. The simulation's initial
conditions and its computational grid are similar to those adopted in
\citet{moscibrodzka:2009} (and references therein).  The initial
condition is a weakly magnetized torus in orbit around a Kerr black
hole with $a_*=0.94$, where $a_*$ is the BH angular momentum.
The inner and outer radius of the initial torus are $6\Rg$ and
$42\Rg$ ($\Rg=\rg$ and $M_{BH}\equiv$ BH mass), respectively.
The difference between the previous \citep{moscibrodzka:2009} and the
current model is the size of the computational domain. Here we extended
the outer boundary to $r = 1000 \Rg$. This large computational
domain is necessary to model the radio spectrum from $1-10^4$ GHz. The
simulation was evolved until $t_{final}=4000 \Rg/c$, which corresponds
to 16 orbital rotations at pressure maximum of the initial disk.

As the simulation advances in time, the magnetorotational instability (MRI)
turns the smooth torus into a turbulent accretion flow. The turbulence
transports angular momentum of the gas outward and inner portions of the disk
fall toward the BH horizon. During the simulation, the accretion flow
produces a variety of outflows. A low-density strongly magnetized
relativistic outflow develops above the BH poles
(\citealt{blandford:1977,mckinney:2004}), the inner accretion disk launches a
mildly relativistic outflow, and subrelativistic winds are produced by the
outer regions of the disk. The latter outflow is formed because the
outer-disk plasma gains an excess angular momentum from the accreting matter. 

\begin{figure*}[htb]
\includegraphics[angle=-90,scale=0.85]{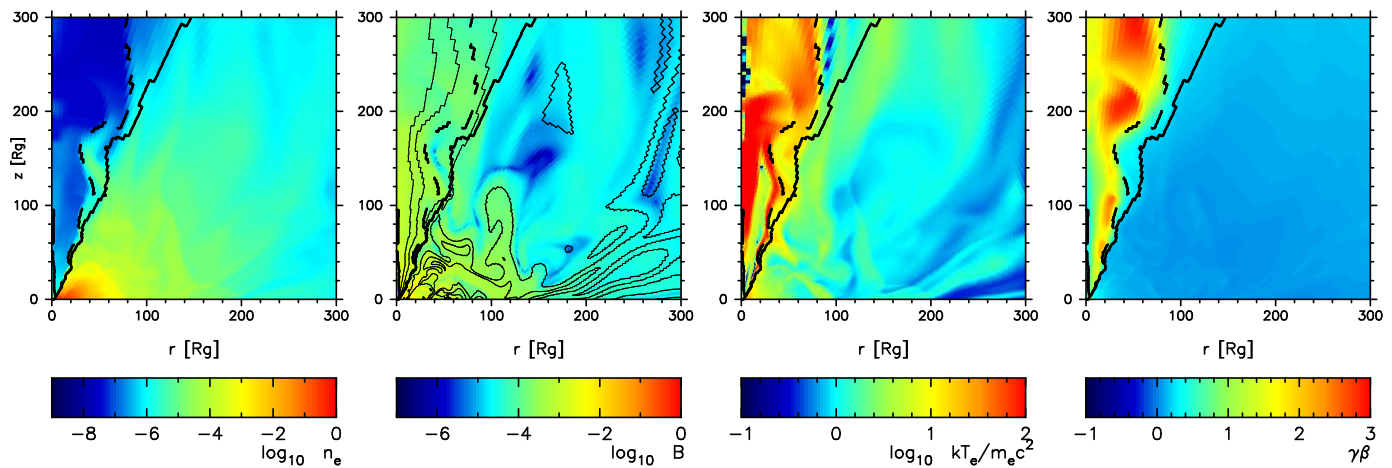}
\caption{
Overall structure of the flow (upper half of the
simulation domain). Panels from left to right show maps of the density, magnetic
field strength, gas temperature, and bulk speed of the gas.  The first
two are given in
dimensionless numerical code units.  Solid and dashed contours illustrate our
jet and disk definitions, which are introduced in Sect.~\ref{model}. The solid line
separates the jet from the disk, and the dashed line separates the jet spine
 from jet sheath.}\label{fig:maps}
\end{figure*}

Fig.~\ref{fig:maps} shows the overall structure of the GRMHD
jet--disk--BH model.  We defined the jet produced by the
accretion flow as an {\em unbound} gas outflowing with a minimum bulk
velocity $\beta_{min}=0.2$~\footnote{$\beta $ is the velocity measured in
a normal observer frame.}. Everything beyond the jet region we
refer to as an accretion disk/flow/wind. In Fig.~\ref{fig:maps}, the
jet region is separated from the accretion flow by a solid contour.
Our formal definition of the jet indicates that it has two
components. The first component is a strongly magnetized nearly empty
funnel where $B^2/\rho_0 > 0.1$, hereafter called jet spine
(the region within the dashed contour  in
Fig.~\ref{fig:maps}).  The second
component is the mildly relativistic outflow along the funnel
  wall (the region between the dashed and solid contours in
  Fig.~\ref{fig:maps}), hereafter called jet sheath. The choice of
$\beta_{min}=0.2$ was arbitrary, but proved to be a relatively robust
threshold to separate jet and disk winds, which exhibit a sharp
boundary (in the jet sheath $\left<\beta\right>=0.4$). With
$\beta_{min}=0.2$ the simulation grid resolves the jet sheath with a
few points. The mass-outflow rate in the jet sheath can be as high as
20\% of the average mass-inflow rate onto the BH. Hence,
radiation from the jet region is dominated by the jet sheath.

The dynamical models of the accreting BH are scale-free, but
the radiative transfer models are not.  We rescaled the numerical model
to the Sgr~A* system.  We fixed the BH mass ($M_{\mathrm BH}=4.5
\times 10^6 {\rm M_{\sun}}$, \citealt{ghez:2008}) and distance (D =
8.4 kpc, \citealt{gillessen:2009}). The remaining model parameters
are the source inclination $i$, the density scaling constant $n_{e,0}$,
and the electron temperature $T_e$. The magnetic-field strength is
given by multiplying the dimensionless $B$ field by $B_{\rm 0} = c
\sqrt{4 \pi m_p n_{e,0}}$. We chose $n_{\rm e,0}$ so that the radio
flux (at $\nu=1-100{\rm GHz}$) matched the observed data points. 
 $\Te$ was parameterized as follows. In the disk we have
  $(\trat)_d = const$, where $\Tp$ is given by the dynamical model and
  $(\trat)_d$ is a free parameter. In the jet we have $
  T_{e,j} = 30 m_e c^2/k$, independently of the proton temperature.
The last assumption was motivated by results presented in the next
section and models of isothermal jets. However, as explained in the
next section, the value of $T_{\rm e,j}$ is similar to the average
$\Tp$ found in the GRMHD model in the jet region, that is,
$(\trat)_j\sim1$.

Spectral energy distributions (SEDs) 
and images of the plasma around the BH are produced by a 3D
general relativistic ray--tracing radiative-transfer code. In the
radiative-transfer 
computations, we assumed that the plasma has a Maxwellian energy
distribution. We used an independent numerical radiative-transfer scheme as
described for instance in \citet{noble:2007}.

\section{Results}~\label{results}

\begin{figure}[htb]
\includegraphics[angle=-90,scale=0.23]{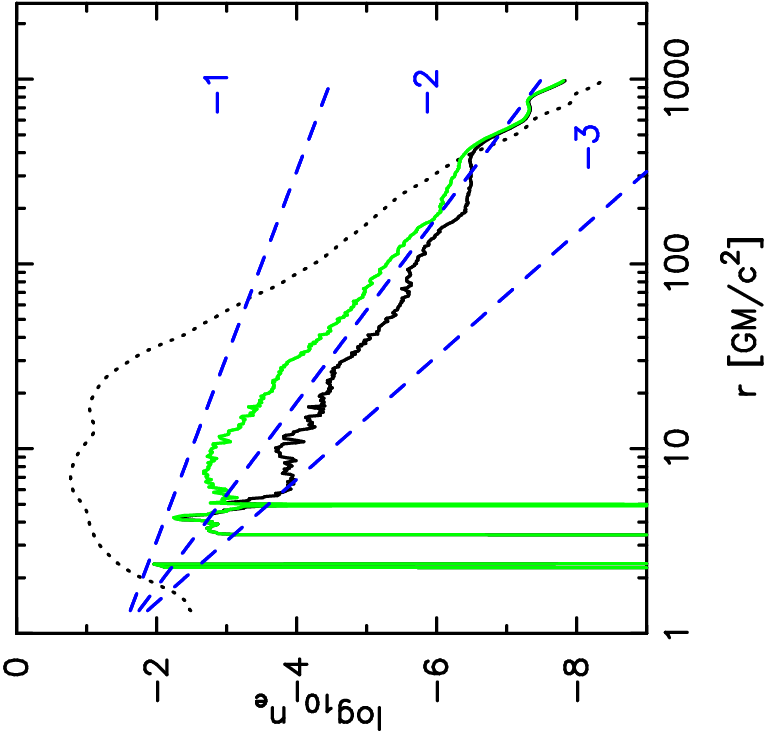}
\includegraphics[angle=-90,scale=0.23]{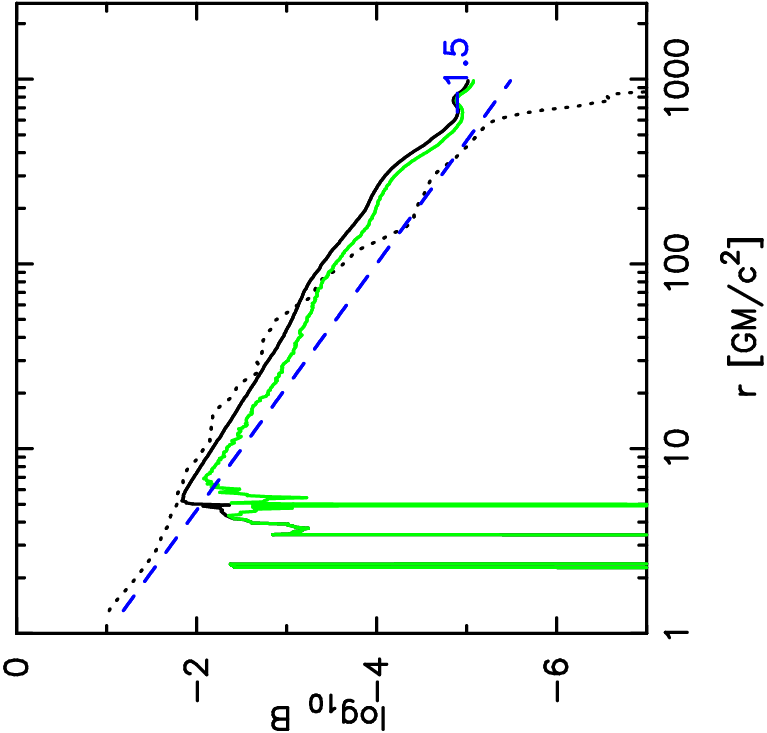}
\includegraphics[angle=-90,scale=0.23]{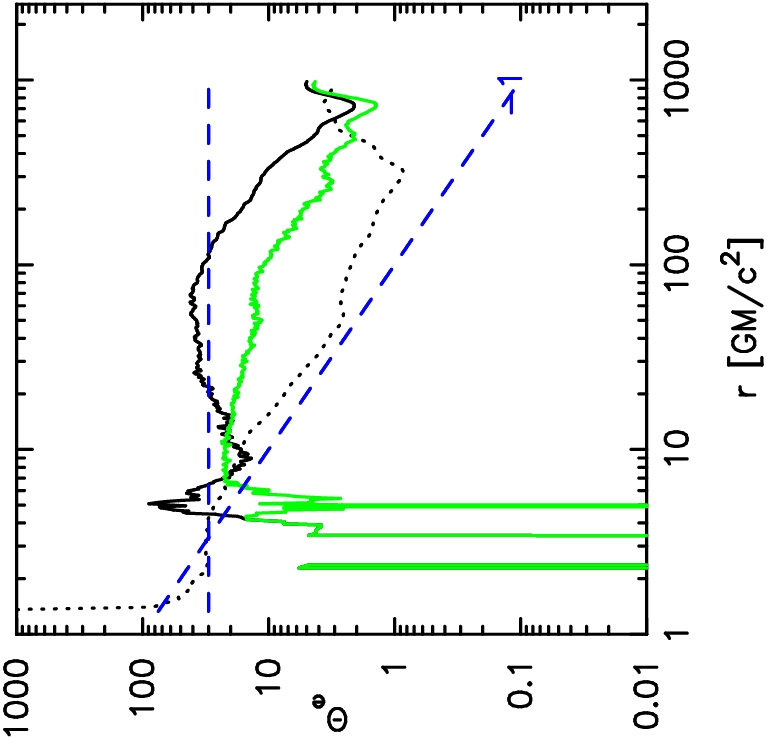}
\caption{
Time- and $\theta$-angle-averaged (see Eq.~\ref{eq:avg}) profiles of $n_{\rm e}$,
$B$, and $\Theta_e=k\Te /m_e c^2$ (assuming $T_p/T_e=1$) in the jet and in the
disk. The dotted lines correspond to the quantities measured in the turbulent
accretion flow, and the solid lines are the averaged profiles of quantities
measured in the jet region.  The solid green lines are the profiles measured
in the jet without the empty jet spine (i.e. excluding zones where
$B^2/\rho_{\rm 0} > 0.1$).  The plasma density and magnetic-field strength are shown
in the numerical code units.} \label{fig:prof}
\end{figure}

We analyzed the radial structure of inflows and outflows produced in the
simulation. 
Fig.~\ref{fig:prof} shows profiles of three quantities measured
in the accretion flow (dotted lines), in the jet spine (solid black lines), and in
the jet sheath (solid green lines). The time- and shell- volume-averaged
radial profiles of $n_{\rm e}$, $B$ and the dimensionless electron temperature,
$\Theta_{\rm e}=k\Te /m_ec^2$, were calculated using the following definition:
\begin{equation}
\left< q(r)\right> = \frac{1}{\Delta t}\int_{t_{min}}^{t_{max}} 
\frac{\int_0^{2\pi}  \int_0^{\pi}  q(r, \theta, \phi , t)\sqrt{-g} d\theta d\phi }
      {\int_0^{2\pi} \int_0^{\pi} \sqrt{-g} d\theta d\phi } dt, \label{eq:avg}
\end{equation}
where $q$ is a scalar quantity, $\sqrt{-g}$ is the determinant of the metric,
$t_{min}=3500 \mathrm{M}$, and $\Delta t=t_{max}-t_{min}=500 \mathrm{M}$.

The radial profiles of the three quantities show approximately power-law
shapes. The density in the accretion flow decreases with radius as $n_{\rm e} \sim
r^{-3}$. The steep power-law dependence is an artifact caused by the adopted
initial conditions (small size of the torus).  The density in the jet
decreases with radius as $n_{\rm e} \sim r^{-2}$.  Close to the BH, the
density of the disk is about 100 times higher than the density in the
outflow. The magnetic fields, both in the inflow and in the outflow, decrease
with distance as $B\sim r^{-3/2}$.  The protons in the accretion flow are near
their virial temperature, $\Tp \sim r^{-1}$. The temperature of the gas in the
jet is approximately constant between $5\Rg$ (where the jet starts) and $r\sim
100\Rg$, and decreases for $r>100\Rg$.  For $r \le 100\Rg$, the gas
temperature is $\left<\Theta_e\right>=30$.  The break in the temperature
power-law at $r\approx 100 \Rg$ may be caused by decollimation of the jet
(because of the small disk size) and/or by a poor numerical resolution of the
model at large radii.

Interestingly, density $n_{\rm e}$ and $B$-fields in the outflow decrease with
radius in a similar manner as the $n_{\rm e}$ and $B$ fields in the semi-analytical
relativistic jet models by \citet{falcke:2000}. Their model, which produces a
flat radio SED, assumes that the temperature of electrons (or, to be more
precise, the electron energy distribution function) is almost constant along the
jet. Hence, we simply adopted an isothermal
jet with $\Theta_e=30$,
in our radiative-transfer models, which corresponds to 
$(\trat)_j\sim1$ up to $r\la 100 \Rg$. Outside the jet $\Te$ is
given by $(\trat)_d=const>1$.

\begin{figure}
\includegraphics[scale=0.4,angle=-90]{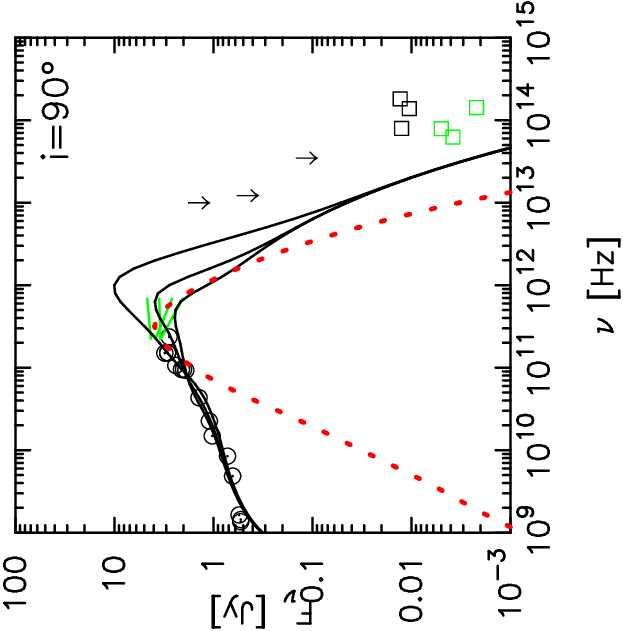}
\includegraphics[scale=0.4,angle=-90]{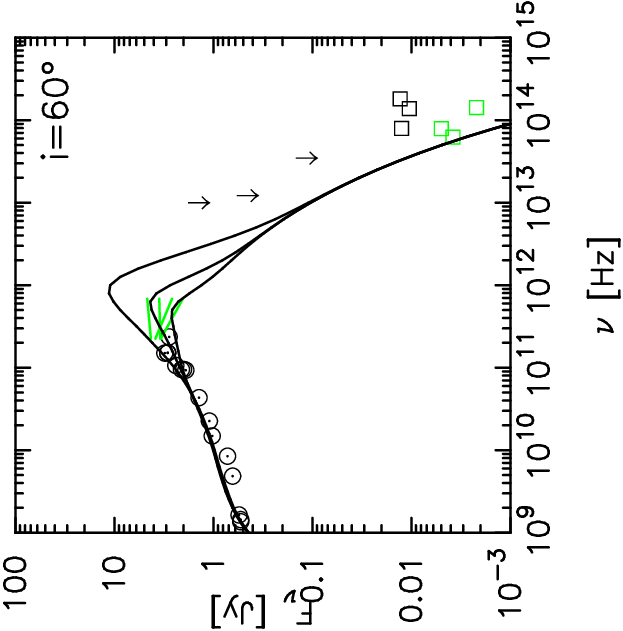}
\includegraphics[scale=0.4,angle=-90]{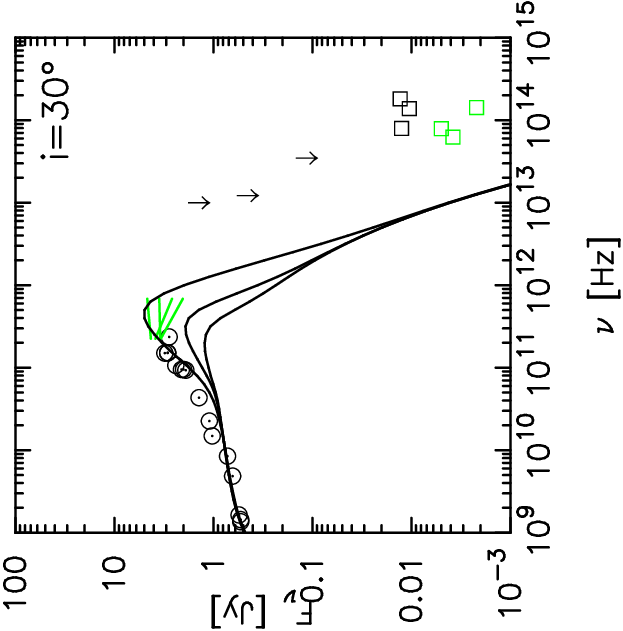}
\includegraphics[scale=0.41,angle=-90]{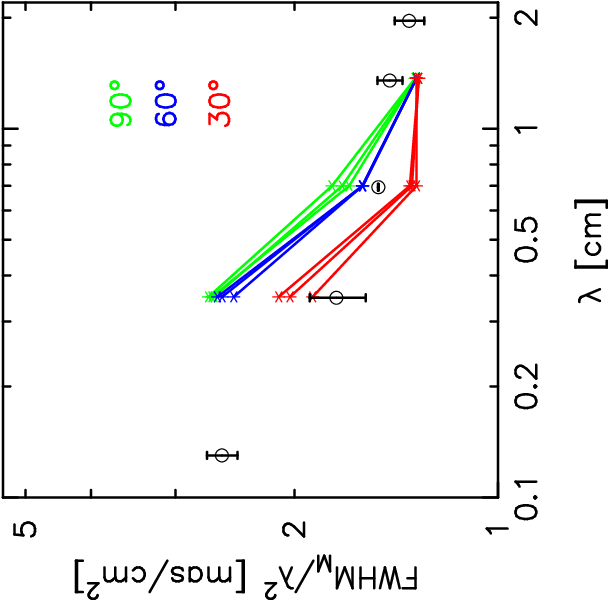}
\caption{
SEDs computed for parameters given in table~\ref{tab:1}.  SEDs are time
averaged. Each line corresponds to a model with a  different temperature ratio in
the accretion disk, $(\trat)_{d}=10,15$ and 20), and various inclinations.  The
red dashed line is the 'best-bet' model from \citet{moscibrodzka:2009}, which
assumes $i\approx90\degr$ and $\trat=3$ in the entire computational
domain. The data points are the same as those used in
\citet{moscibrodzka:2009}. The bottom-right panel shows the sizes of the GRMHD jet
model {\em including scatter-broadening}.  Open symbols are Sgr~A* measured sizes from
\citet{bower:2006}.}\label{fig:sed}
\end{figure}

The radiative-transfer model parameters are given in Table~\ref{tab:1}. 
Fig.~\ref{fig:sed} shows time-averaged SEDs of the jet-plus-disk system. In Fig.~\ref{fig:sed}, three
lines in each panel correspond to models with various $(\trat)_d=10,15$, and $20$.

Indeed, the jet spectrum is nearly flat, whereas the disk produces a hump at
submillimeter wavelengths. The two components together are able to reproduce
the spectrum remarkably well. At this point, we do not intend to conduct a full
statistical exploration of the parameter space and, therefore, there is no strong
reason to favor any specific parameter combination yet. However, it is already
clear that the jet is crucial for filling in the low-frequency radio spectrum.

As predicted in \citet{yuan:2002}, the disk $\Te$ has to be low and $(\trat)_d
\sim 10-20$ to smoothly connect jet and disk emission. The spectral
shape at lower frequencies depends weakly on the observing inclination angle.
For $i=30\degr$, the slope of the SED is somewhat shallower
($\alpha_{\nu}\sim0$) than the slope seen under higher
inclinations ($\alpha_{\nu}\sim 0.3$). This dependence is in line with
previous analytical estimates \citep{falcke:1999}.

Are size and structure of Sgr~A* also consistent with the model?  Scattering
is, of course, a major problem and the relative orientation between scattering
disk and jet axis adds another free parameter. We approached this here by
scatter-broadening our model images for one arbitrary orientation and
comparing them to the available data \citep{falcke:2009}. Given the very
limited structural information we have -- a size in east-west direction --
this is the best we can do for now.

We proceeded as follows: Our radiative-transfer model computes images of the
flow at $\lambda=3,7$ and $13$ millimeters. The synthetic images are then
convolved with the symmetric scattering Gaussian profile with a $FWHM=1.309
\lambda^2$ \citep{bower:2006}. Next, the size of the scatter-broadened image
is measured by computing the eigenvalues of the matrix formed by taking the
second angular moments of the image (the principal axis lengths). This method
yields an accurate size of the emitting region if the brightness distribution
is Gaussian-like and is therefore most accurate for strongly scattered images.

In the bottom-right panel of Fig.~\ref{fig:sed}, we compare the measured
major-axis 
sizes of Sgr~A* at $\lambda=0.3-2\rm{cm}$ with those predicted by the
model. We did not extend the model size to $\lambda=1.3mm$, because here the
simulated image is highly non-Gaussian and a single major-axis size is
meaningless. Furthermore the VLBI data were measured on a single, very narrow triangle
of baselines only. We also cannot extend our model sizes to $\lambda >
2\rm{cm}$ because the size of the scattered image becomes larger than our
computational domain.  However, in the region around $\lambda \sim
0.7\rm{cm}$, where the sizes are most reliably determined, the major-axis
sizes are of the correct order. Moreover, the images look Gaussian-like despite
the underlying jet-structure.

\section{Summary}\label{discussion}

We have reanalyzed the structure of inflows (disk) and outflows (jets)
produced in some GRMHD models of accreting BH \citep{gammie:2003}.  We
pointed out that the time-averaged radial profiles of plasma density and
B-fields in the jet-like outflows are similar, though not exactly identical,
to those used in semi-analytic jet models that readily explain the radio
SED of Sgr~A* \citep{falcke:2000,yuan:2002}
based on a scaled-down \citet{blandford:1979} model combined with a radiatively
inefficient accretion flow or RIAF.  Consequently, we were able for the first
time to relatively easily reproduce the radio spectrum and size of Sgr A* at
mm-waves with a standard GRMHD model by simply allowing for different
electron heating in jet and disk.

In particular, we found that the flat radio spectrum of jets is indeed
reproduced by the simulations when we kept the electron temperature
constant along the jet, as assumed in the \citet{blandford:1979}
model. In the inner parts the jet is well-described by a single-temperature 
plasma with a proton-to-electron temperature ratio on
the order unity. To reproduce the overall spectrum, however, the electron
temperature in the disk needs to be lower than in the jet by at least
an order of magnitude to avoid strong absorption and huge emission at
submillimeter waves.  In turn, this requires the presence of a
two-temperature plasma in the accretion flow, with a high
proton-to-electron temperature ratio $(\sim10-20)$. This is expected because 
two-temperature plasmas were actually postulated in the earliest models for RIAFs
\citep{narayan:1995}. Interestingly, $\trat=10-20$ is consistent
  with results of local (shearing box) simulations of collisionless
  plasma in which kinetic effects are included \citep{sharma:2007}.

 In the jet models, $\mdot$ is about 20 times higher than 
  $\mdot\approx2\times10^{-9} \mdotu$ in the '$\trat=3$ everywhere'
  best-bet model from \citet{moscibrodzka:2009}.  Our
  $\mdot=4.5\times10^{-8} \mdotu$ (see Table 1) is consistent with
  $\mdot=6\times10^{-8} \mdotu$ found in models where the BH is fed by
  stellar winds (\citealt{roman:2010}, but see
  \citealt{quataert:2004}), and with estimates by
  \citet{sharma:2007} based on local collisionless plasma models. 
A higher $\dot{M}$ may also change
expectations for the recently discovered cloud-like object G2 that moves
toward Sgr~A* \citep{gillessen:2012}, which is expected to be accreted
onto the BH soon. For a higher pre-impact accretion rate the increase
in $\mdot$ due to the cloud would be much less dramatic than predicted
(e.g., in \citealt{moscibrodzka:2012} or \citealt{sadowski:2013}).

 Summarizing, the electron distribution function in GRMHD models
  is a free function that can vary with space and time. We showed that
  a {\it natural} modification of this distribution function produces 
  SEDs that fit the observations well.  The
re-heating (or re-acceleration) of electrons in the jet might be due
to effects that were described in the GRMHD models, such as shear,
strong magnetization, or shocks in the jet sheath, but this requires a more detailed
investigation.  We conclude that the exact nature of the electron
heating in jets and disks deserves more attention in the future.

\begin{acknowledgements} 
We thank C. Gammie and J. Dexter for comments.
\end{acknowledgements}


\end{document}